\documentstyle[aps,prl]{revtex}

\begin{document}
\twocolumn[
{\hsize\textwidth\columnwidth\hsize\csname@twocolumnfalse\endcsname
\draft
\title{Evidence of quantum criticality in the doped Haldane system
Y$_2$BaNiO$_5$}
\author{C. Payen, E. Janod, K. Schoumacker}
\address{Institut des Materiaux Jean Rouxel,Universite de Nantes-CNRS,\\
44322 Nantes\\
Cedex 3, France}
\author{C. D. Batista, K. Hallberg, A. A. Aligia}
\address{Centro At\'{o}mico Bariloche and Instituto Balseiro,\\
Comisi\'{o}n Nacional de Energ\'{\i}a At\'{o}mica 8400 S.C. de Bariloche,\\
Argentina}
\date{\today}
\maketitle

\begin{abstract}
Experimental bulk susceptibility $\chi (T)$ and magnetization $M(H,T)$ of
the S=1 Haldane chain system doped with nonmagnetic impurities, Y$_{2}$BaNi$%
_{1-x}$Zn$_{x}$O$_{5}$ ($x\le 0.08$), are analyzed. A numerical calculation
for the low-energy spectrum of non-interacting open segments describes very
well experimental data above 4 K. Below 4 K, we observe power-law behaviors,
$\chi (T)$$\sim $ $T$$^{-\alpha }$ and $M(H,T)\sim $ $T$$^{1-\alpha }$$f$$%
_{\alpha }$$(H/T)$, with $\alpha $ $(<1)$ depending on the doping
concentration $x$. This observation suggests the appearance of a gapless
quantum phase due to a broad distribution of effective couplings between the
dilution-induced moments.
\end{abstract}

\pacs{75.10.Jm, 75.45.+j, 75.40.Cx, 75.40.Mg}
}
]
The past two decades have seen a resurgence of interest devoted to
macroscopic quantum phenomena in Heisenberg antiferromagnets (HAF). A
current issue in this field concerns the interplay between quantum spin
fluctuations and quenched (i.e., time-independent) disorder in one dimension
(1D) \cite{1}. A simple nontrivial model to address this issue is the 1D-HAF
with spins $S=1$, which has a Haldane gap $\Delta \approx 0.4J$ and a short
correlation length $\xi \approx 6$ in its spin-liquid ground state ($J$ is
the nearest-neighbor exchange parameter) \cite{2,3}. In this system, the
disorder in the form of site depletion leads to interesting quantum
phenomena due to the creation of two effective $S=1$/2 spins on the opposite
edges of a segment, near the vacancies \cite{4,5}. The inclusion of weak
magnetic bonds across the spin-vacancies induces bond disorder and causes
departure from simple finite-size behaviors. In this case, some recent
theoretical works, for the $S=1$ chain with variable nonmagnetic doping,
found a quantum (zero-temperature) phase transition from the Haldane phase
to a 1D random-singlet (RS) phase in which effective spins are coupled into
singlets over all length scales \cite{6,7}. On the experimental side,
however, this problem has been very little explored from the point of view
of quantum criticality. While it is known that some $S=1$ diluted compounds
can sustain paramagnetism down to low temperature, most of the existing
experimental studies aimed to confirm the existence of $S=1/2$ end-chain
states or to reveal the internal structure of these end states \cite{4,8}. A
noticeable exception is the recent observation of a 3D long-range order
(LRO) caused by nonmagnetic doping in PbNi$_2$V$_2$O$_8$ \cite{9}. This
Haldane system shows, however, substantial interchain interactions, so it is
almost critical towards the formation of a LRO in the absence of intentional
disorder.

In this work, we examine the effects of nonmagnetic doping on the bulk
magnetic responses in the $S=1$ quasi-1D-HAF Y$_{2}$BaNiO$_{5}$. We chose
this well characterized system because of the rather high value of its main
energy scale, $J$$\approx $280 K, and the very small value of the interchain
coupling, $|J_{\perp }/J|<10^{-3}$ \cite{10}. Furthermore, it has a simple
orthorhombic crystal structure \cite{11} and a small single-ion anisotropy ($%
D/J$$\approx $-0.04, $E/J$$\approx $-0.01) \cite{10}. A weak nonmagnetic
doping in Y$_{2}$BaNiO$_{5}$, which is realized by substituting Zn$^{2+}$ or
Mg$^{2+}$ ions for the $S=1$ Ni$^{2+}$ ions, yields new magnetic states
below the Haldane gap ($\Delta $$\approx $100 K) with no sign of LRO down to
very low temperature \cite{12,13,14}. Recently, the low-temperature specific
heat of Y$_{2}$BaNi$_{1-x}$Zn$_{x}$O$_{5}$ ($x=0.04$) and the ESR spectra of
a series of Mg-doped compounds have been quantitatively explained by an
effective model which describes the low-energy spectrum of non-interacting
open $S=1$ chains \cite{15,16}. These results and the recent NMR experiments
in Ref. 8 are strong indications of the existence of $S=1/2$ end-chain
states. In the present work, we analyze new susceptibility and magnetization
data for Y$_{2}$BaNi$_{1-x}$Zn$_{x}$O$_{5}$ with 0.04$\leq $x$\leq $0.08.
Above 4 K, the simple model of non-interacting segments used in Refs. \cite
{15,16} describes the data very well. However, at very low temperature we
observe a new regime with divergent power law behavior, which is similar to
that observed in some doped semiconductors \cite{17}. Our results support
the existence of a zero-temperature gapless phase due to randomness in the
effective bond distribution that is reinforced by the interchain
interactions. The low-energy effective model for this phase resembles a 2D
or 3D RS state.

Figure 1 shows the linear susceptibilities, $\chi (T)$, measured with a
commercial SQUID magnetometer for a series of polycrystalline samples of Y$%
_{2}$BaNi$_{1-x}$Zn$_{x}$O$_{5}$ ($x=0$, 0.04, 0.06 and 0.08) that were
prepared through standard solid-state reactions \cite{11}. The results
obtained for the nominally pure sample compare well with those in the
published literature \cite{12}. At low temperatures, $\chi (T)$ is well
fitted by the sum of a $T$-independent term $\chi $$_{0}$, a Curie law and a
thermally activated term with $\Delta $$\approx $100 K (see Fig.1). The sum
of the activated part and the constant $\chi _{0}\approx 10^{-4}$ cm$^{3}$%
/mol is negligible below 15 K. The low-temperature susceptibility (below
10-15 K) for $x=0$ is mainly due to the Curie contribution that corresponds
to 1.4 percent of free $S=1/2$ spins per formula unit. This Curie behavior
can be interpreted as coming from natural crystal defects and a weak excess
oxygen. The main effect of the intentional nonmagnetic dilution is to
increase drastically the low-temperature upturn in $\chi (T)$, as is evident
from Fig. 1. This indicates the existence of paramagnetic moments whose
amount is related to the doping concentration. The behavior of $\chi (T)$ is
however qualitatively the same, regardless of the level of dilution. The
data for the doped compounds were first analyzed with a modified Curie-Weiss
law, $\chi (T)=\chi _{0}+C/(T-\theta )$, between 1.9 and 15 K. A small
phenomenological $\theta $$\approx $-0.6 K was required to describe the
data. The dependence of the Curie constant $C$ as a function of the Zn
doping is shown in the inset of Fig. 1. At low doping, the observed $C$
value is that for two $S=1/2$ spins per nonmagnetic impurity. At high $x$,
however, the Curie constant should be that for one half of a spin-1 per Zn
since a short segment with an odd (even) number of spins behaves as a $S=1$ (%
$S=0$) 'molecular' entity at low temperature and the number of odd chains is
half the total number of chains. This tendency seems to be observed (see
Fig. 1). We have also measured the magnetization, $M(H,T)$, for applied
field up to 5 T. For all samples, $M(H,T)$-$\chi $$_{0}$$H$ shows a
non-Brillouin behavior and is smaller than twice the Brillouin function for
one $S=1/2$ spin per Zn.

To go further in our analysis, we have compared the experimental data with a
theoretical model based on the picture of non-interacting segments, as
described hereafter. The energy spectrum of finite segments described by the
$S=1$ Heisenberg Hamiltonian is characterized by four low-energy states, a
singlet and a triplet \cite{3}. It corresponds to the existence of a
coupling between the two $S=1/2$ edge spins, which may be AF or
ferromagnetic (F). The energy difference among these levels decreases
exponentially with the length of the chain $N$, as $(-1)^{N}e^{-N/\xi }$,
and already for chains of a few lattice sites, these four states are
separated from the rest of the spectrum by an energy of the order of the
Haldane gap \cite{3}. Therefore at temperature $T$ and magnetic energies $%
\mu _{B}H$ much lower than the Haldane gap, the magnetic and thermal
properties of a system composed of segments of several lengths are
completely described by the energy of these four states and the matrix
elements of the magnetization operator in this reduced subspace. We have
calculated these quantities using the Density Matrix Renormalization Group
(DMRG) method. The resulting Hamiltonian which describes the low-energy
properties of a segment including the triplet $\left|
1,S_{z}\right\rangle $ and the singlet state $\left| 0\right\rangle $ is:

\begin{eqnarray}
H_{eff} &=&E_{0}(N)+(J\alpha (N)+D\beta (N))\left| 0\right\rangle
\left\langle 0\right| +D\gamma (N)S_{z}^{2}  \nonumber \\
&&+E\gamma (N)(S_{x}^{2}-S_{y}^{2})-\mu _{B}\sum_{\nu \alpha }H^{\alpha
}g^{\alpha \nu }S_{t}^{\nu },  \label{ecu2}
\end{eqnarray}
where $E_{0}(N),\;\alpha (N)$, $\beta (N)\;$and $\gamma (N)$ are functions
of the chain length $N$ (determined from the DMRG data). The validity of the
last term has been verified explicitly by calculating the matrix elements of
$S_{t}^{+}$, and $S_{t}^{-}$ for all chains. The magnetic moment operator is
-$\nabla _{H}H_{eff}$, and to compare with experiments in polycrystalline
samples, we have averaged its expectation value over all possible
orientations of the crystal. Also, a distribution of chain segments
corresponding to a random distribution of defects was assumed \cite{16}.
Figure 2 shows the comparison between the measured $\chi (T)T$ products for $%
T<12$ K and the numerical solution. The adjustable parameters in the
calculations were the doping $x$ and $\chi $$_{0}$. All the other parameters
were held fixed at the values deduced from previous works \cite{10,16}: $%
J=280$ K, $D/J=-0.039$, $E/J=-0.013$, $g_{x}=g_{y}=2.17$ and $g_{z}=2.20$.
The fitted values of $x$ ($x=0.041$, 0.060 and 0.074) are close to the
nominal concentrations, and the fitted $\chi $$_{0}$ remains in the expected
range ($\chi _{0}=1.0\times 10^{-4}$ to $1.9\times 10^{-4}$ cm$^{3}$%
/Ba-mol). Taking into account that uncertainties in the above parameters
might affect the theoretical curve, the success of the anisotropic model of
non-interacting segments to explain the data above 4 K is remarkable. Note
that for the cases with nominal concentration of Zn $x=0.06$ and $x=0.08$, a
decrease of the order of 5\% or less in the actual $x$ used, would improve
considerably the fitting above 5 K, but increase the differences between
theory and experiment in the low-temperature part. Instead, because of the
different rate of change of both curves, it is not possible to obtain a good
fit for $T<4$ K by small changes of $x$ or other parameters. A good
agreement between the
experimental magnetization ($M(H,T)$-$\chi $$_{0}$$H$) and the numerical
data above 4 K (not shown) is also achieved using the parameters required to
describe the susceptibility data. This noticeable agreement confirms the
existence of the $S=1/2$ end-chain excitations. Note that an accurate
description of the experimental data needs to consider both, a random
distribution of defects \cite{16} and the effect of the
single-ion anisotropy \cite{15,16}, which is stronger for shorter segments.

However, below 4 K, the experimental data deviates from the predictions of
the above described model. This is seen in Fig. 2 for the susceptibility. It
is also observed for our magnetization data $M(H,T)$-$\chi $$_{0}$$H$. In
previous studies of specific heat and electron spin resonance, while
excellent agreement was obtained above 4 K, a discrepancy was also found
below this temperature and interpreted as the onset of interchain
interactions not taken into account in this model \cite{15,16}. An analysis
of the susceptibility shows that, in the vicinity of 4 K, there is a
crossover to a sub-Curie power law regime, $[\chi (T)-\chi _{0}]\sim
T^{-\alpha }$ with $\alpha =0.83$, 0.79 and 0.76 for $x=0.04$, 0.06 and
0.08, respectively (see Fig. 3). The susceptibility data for the Mg-doped
compound Y$_{2}$BaNi$_{0.959}$Mg$_{0.041}$O$_{5}$ in Ref. \cite{14} also
obey a power-law form with $\alpha =0.73$ (see Fig. 3). This behavior,
which cannot be reproduced by the model of non-interacting segments, is
reminiscent of some random exchange Heisenberg systems for which the AF
couplings between nearest-neighbor $S=1/2$ spins are random in their
magnitude, such as insulating phosphorous-doped silicon (Si:P) \cite{17}.
Bhatt and Lee have proposed a theoretical method to explain the properties
of Si:P \cite{18}. Within this approach, the spin pairs with the strongest
AF bonds freeze into inert singlets, leaving behind a new ensemble of active
spins with a renormalized distribution of exchange, and the quantum
fluctuations drive the whole system into a ground state consisting of random
local singlets (RS phase). The behavior of the susceptibility, $\chi (T)$$%
\sim $$T$$^{-\alpha }$ with $\alpha <1$, follows from a divergent power-law
distribution of the renormalized exchange, $P(J)$$\sim $$J$$^{-\alpha }$. A
scaling of the magnetization has been also predicted and experimentally
observed \cite{17}, $M(H,T)/T^{1-\alpha }=f_{\alpha }(H/T)$. A similar
scaling is, in fact, very well obeyed by our data below (but not above) 4 K
using the $\alpha $ exponents determined from the susceptibility. Fig. 3
shows a plot of $(M(H,T)-\chi _{0}H)/T^{1-\alpha }$ versus $H/T$ for each of
the three samples studied. All the data for each sample, for different
magnetic fields and temperatures, lie on a single scaling curve, which is
however different from the expression of $f_{\alpha }(H/T)$ calculated on
the basis of the Bhatt and Lee's solution for Si:P \cite{17}. This is
illustrated in Fig. 3 for Y$_{2}$BaNi$_{0.94}$Zn$_{0.06}$O$_{5}$.

These power law behaviors show that the picture of the ideal 1D anisotropic
Heisenberg non-interacting chain breaks down below 4 K, and that additional
interactions play a role. From simple arguments based on perturbation
theory, one expects that the dominant inter-chain interaction is a coupling $%
J$$_{\perp }$$_{b}$ between $S=1$ spins lying in nearest-neighbor chains
along the $b$ direction, perpendicular to the chain axis $a$. A calculation
using the cell perturbation method gives $J$$_{\perp }$$_{b}$$\approx $0.2 K
\cite{19}. This interaction should be at least an order of magnitude larger
than that between two spins on the same atomic chain, with a Zn atom in
between. As discussed in more detail in Ref. \cite{15,19} $J$$_{\perp }$$%
_{b} $ induces an effective interaction $J^{\prime }$ between $S=1/2$
end-states of neighboring chains along the $b$ direction, the magnitude and
sign of which depends on the detailed position of the defects. The maximum
absolute value of $J^{\prime }$ is of the order of $10J_{\perp b}$ and it
should decay exponentially with the distance along the chain between the
defects. Since the interaction between $S=1/2$ defects lying in different $ab
$ planes is expected to be at least an order of magnitude smaller, the
physics in the range 0.1 K$<T<$4 K seems to correspond to a 2D array of $%
S=1/2$ spins, with random F and AF interactions coming from exponentially
decaying intra- and interchain couplings.

In agreement with this, the spin glass behavior observed below 3 K in Ca
doped systems Y$_{2-y}$Ca$_{y}$BaNiO$_{5}$ indicates a dimensionality higher
than one in the system at low temperature \cite{14,19}. For a 2D or 3D $%
S=1/2 $ model with random AF interactions, Bhatt and Lee obtained a power
law behavior with $\alpha $ $(<1)$ depending on the doping concentration
\cite{18}. This result is consistent with our observations but, as mentioned
above, we obtain a different scaling function for the magnetization. This
might be due to the presence of effective F coupling in Y$_{2}$BaNi$_{1-x}$Zn%
$_{x}$O$_{5}$. Our results can also be compared to a recent theoretical work
in which the spin-1 chain with an AF coupling across nonmagnetic impurities
is mapped onto a bond disordered spin-1/2 chain \cite{6}. Within this 1D
approach, the Haldane state is found to be stable to weak dilution. For
sufficiently strong randomness, however, the system flows towards a 1D RS
phase with diverging correlation length. In this gapless phase, the form of
the low energy asymptotic behavior of $\chi (T)$ is independent of the level
of the disorder, $\chi (T)\sim 1/(T\ln ^{2}T)$. In the Haldane phase, there
is a Griffiths gapless region where $\chi (T)$ varies as $T$$^{-\alpha }$
with $\alpha $ $(<1)$ which is dependent of the details of the randomness.
It is unclear however whether a similar physics should be valid or not when
interchain coupling is turned on \cite{20}. Furthermore, the form of the
asymptotic behavior of $\chi (T)$ is not known for a 2D or 3D RS phase \cite
{1}. If there were no interchain coupling $J$$_{\perp }$, the low-energy
model for Y$_{2}$BaNi$_{1-x}$Zn$_{x}$O$_{5}$ with $0.04\leq x\leq 0.08$
would be the 1D RS phase depicted in Ref. \cite{6} since the bond across a
Zn is expected to be much weaker than the average coupling between the $%
S=1/2 $ edges within a segment \cite{15}. From our observations, it is
tempting to say that the $H/T$ variable in the scaling of $M(H,T)$ indicates
a $T_{c}=0$ critical point and that Y$_{2}$BaNi$_{1-x}$Zn$_{x}$O$_{5}$ has a
quantum critical ground state even for the smallest doping $x=0.04$. From
the theoretical point of view, the detailed explanation of the observed
power-law behaviors remains open.

Interestingly, a similar singular behavior has been disclosed for two $S=1$
1D-HAF compounds, namely AgVP$_{2}$S$_{6}$ and NENP, on nominally pure
samples \cite{20}. A broad distribution of couplings between the
dilution-induced moments and quantum fluctuations drives Y$_{2}$BaNi$_{1-x}$%
Zn$_{x}$O$_{5}$, and perhaps other real systems as well, towards a novel
ground state which resembles a random valence bond state. This is in
contrast with the LRO observed in PbNi$_{2}$V$_{2}$O$_{8}$ \cite{9}.
Therefore there is no universal behavior under nonmagnetic doping in real $%
S=1$ 1D-HAF substances. Details of the topology of the magnetic interaction
lattice are probably crucial in determining the way the Haldane phase is
destabilized by the quenched disorder.

K.H. and C.D.B. are supported by CONICET, Argentina. A.A.A. is partially
supported by CONICET. C.P. is partially supported by the 'Institut
Universitaire de France'. C.P. thanks H. Mutka for helpful discussions and
continued support. K. H. is grateful to the Physics Department of the
University of Buenos Aires for the hospitality during this work. This work
was supported by PICT 03-00121-02153 of ANPCyT and PIP 4952/96 of CONICET.

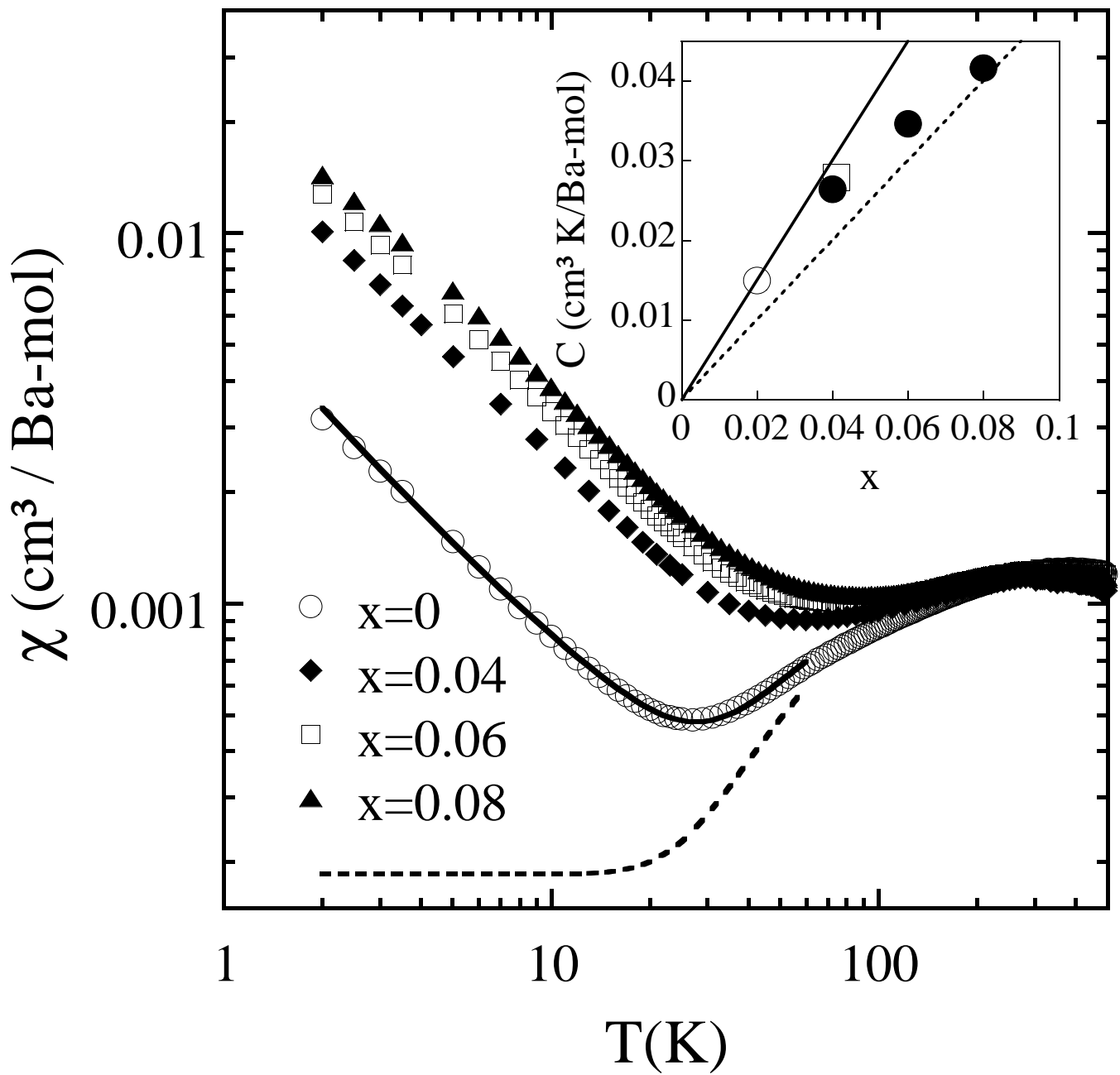
\begin{figure}[tbp]
\caption{Magnetic susceptibility versus temperature, $\chi(T)$, for Y$_2$BaNi%
$_{1-x}$Zn$_x$O$_5$ (0$\leq$x$\leq$0.08). The solid line is a fit for x=0
with $\chi$=$\chi_0$+$C$/($T$-$\theta$)+$A$($\Delta$/T)$^{1/2}$e$%
^{-\Delta/T} $. The sum of the first and third terms is shown as a dashed
line. Inset: the doping dependence of $C$ as described in the text. Data
from Ref. [12] (open circle) and [14] (square) for Mg-doped Y$_2$BaNiO$_5$
are included for comparison. Solid and dashed lines show the result assuming
that one impurity creates two free S=1/2 spins and one half of a S=1 spin,
respectively.}
\label{fig1}
\end{figure}

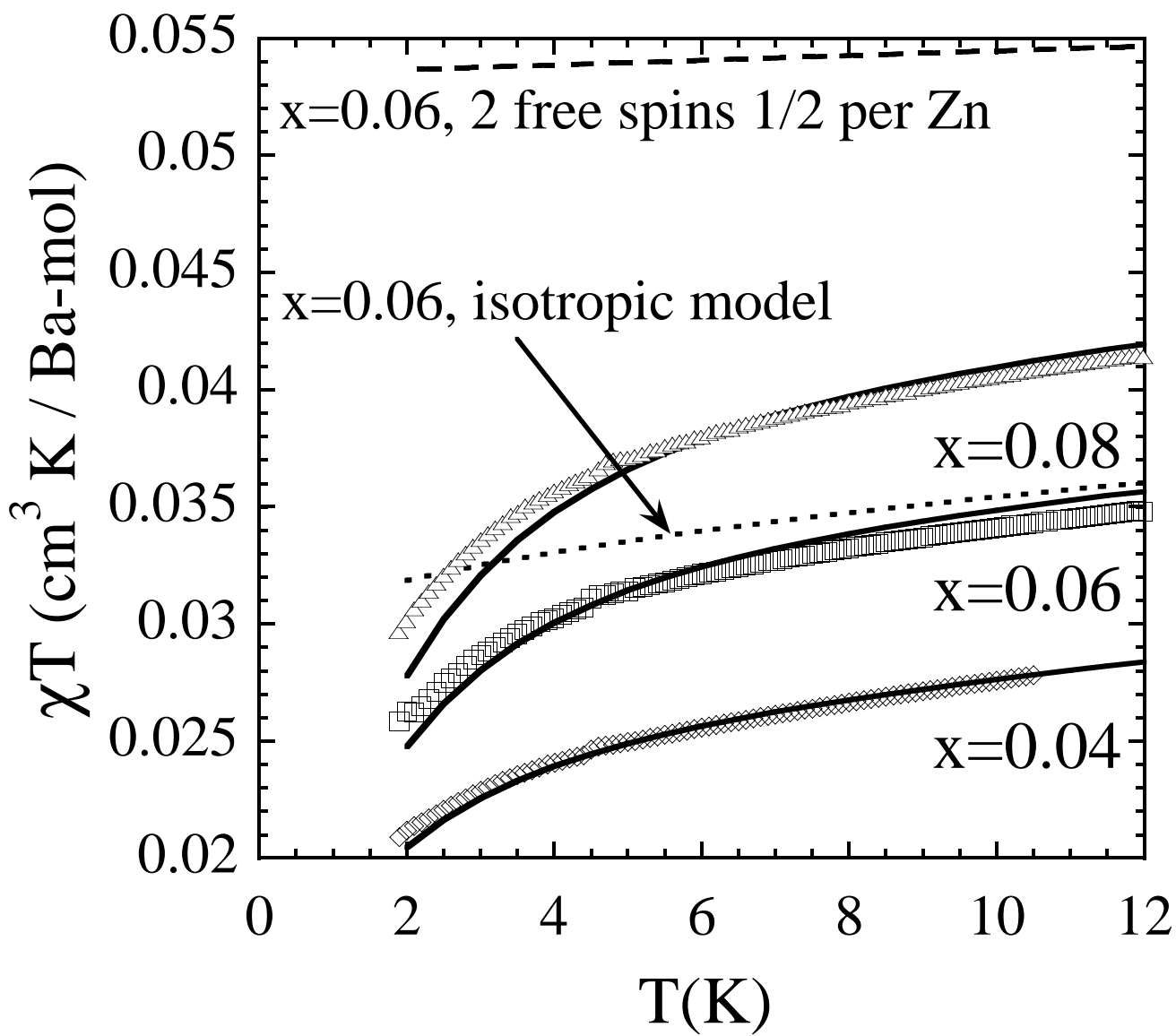
\begin{figure}[tbp]
\caption{$\chi$$T$ versus $T$ plot for Y$_2$BaNi$_{1-x}$Zn$_x$O$_5$ (0.04$%
\leq$x$\leq$0.08) in the range 1.9-12 K. Errors bars are of the order of the
sizes of the symbols. The solid lines show the best fits with our numerical
solution, as described in the text. The dashed and dotted lines are for the
two free S=1/2 spins prediction and a numerical calculation neglecting the
single-ion anisotropy, respectively, for x=0.06.}
\label{fig2}
\end{figure}

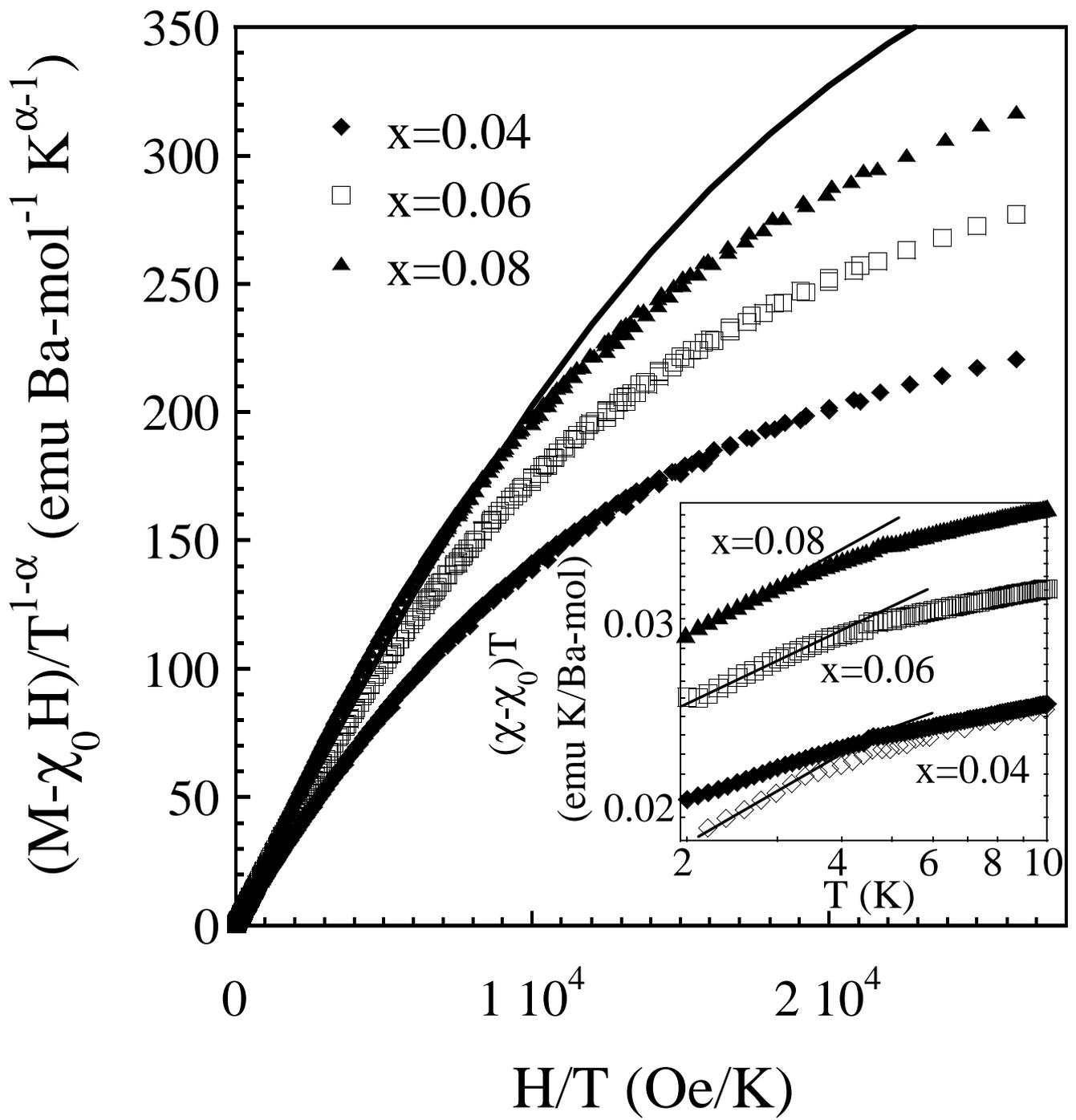
\begin{figure}[tbp]
\caption{Power-law scaling of the magnetization, $[{M(H,T)}$-$\chi_0$]/$T$$%
^{1-\alpha}$=$f$$_{\alpha}$($H/T$), for Y$_2$BaNi$_{1-x}$Zn$_x$O$_5$ (0.04$%
\leq$x$\leq$0.08) in the range 1.9-4 K with data taken every 0.1K. The
values of $\alpha$ are determined from the susceptibilities. The solid line
is the theoretical expression of $f$$_{\alpha}$($H/T$) from Ref. [17] for
x=0.06. Inset: Double-logarithmic plot of [$\chi(T)$-$\chi_0$]$T$ versus $T$
for Y$_2$BaNi$_{1-x}$Zn$_x$O$_5$. Data for Y$_2$BaNi$_{0.959}$Mg$_{0.041}$O$%
_5$ in Ref. [14] are shown as open diamonds.}
\label{fig3}
\end{figure}

\end{document}